\documentclass[authoryear,review,11pt,a4paper]{elsarticle}%
\usepackage{comment}
\usepackage{amsfonts}
\usepackage{amsmath}
\usepackage{mitpress}
\usepackage{amssymb}
\usepackage{graphicx}
\usepackage{graphicx}
\usepackage{lscape}
\usepackage{booktabs}
\usepackage{multirow}
\usepackage{float}
\usepackage{caption}
\usepackage{subcaption}
\usepackage{lineno}
\usepackage{hyperref}
\usepackage{amsmath}
\usepackage{amsfonts}
\usepackage{amssymb}
\usepackage{setspace}
\usepackage{fullpage}
\usepackage{color}
\usepackage{amsmath}
\usepackage{amsfonts}
\usepackage{amssymb}
\usepackage{amsthm}
\usepackage{bbm}
\usepackage{tikz}
\usepackage{tikz-cd}
\usepackage{algorithm}
\usepackage{algpseudocode}%
\providecommand{\U}[1]{\protect\rule{.1in}{.1in}}
\biboptions{authoryear}

\newdefinition{rmk}{Remark}
\newproof{pf}{Proof}
\journal{Econometrics and Statistics}
\newdimen\dummy
\dummy=\oddsidemargin
\begin{document}

\title{Dynamic models using score copula innovations}
\author{Landan Zhang\footnotemark[1], Michael K. Pitt$^{1}$ and Robert
Kohn\footnotemark[2]\\$^{1}$Department of Statistics, King's College London, UK.\\$^{2}$Department of Economics, University of New South Wales, Australia.}

\begin{abstract}
This paper introduces a new class of observation driven dynamic models. The
time evolving parameters are driven by innovations of copula form. The
resulting models can be made strictly stationary and the innovation term is
typically chosen to be Gaussian. The innovations are formed by applying a
copula approach for the conditional score function which has close connections
the existing literature on GAS models. This new method provides a unified
framework for observation-driven models allowing the likelihood to be
explicitly computed using the prediction decomposition. The approach may be
used for multiple lag structures and for multivariate models. Strict
stationarity can be easily imposed upon the models making the invariant
properties simple to ascertain. This property also has advantages for
specifying the initial conditions needed for maximum likelihood estimation.
One step and multi-period forecasting is straight-forward and the forecasting
density is either in closed form or a simple mixture over a univariate
component. The approach is very general and the illustrations focus on
volatility models and duration models. We illustrate the performance of the
modelling approach for both univariate and multivariate volatility models.

Key words: GARCH, heteroskedasticity, GAS models, Copula, Time Series.
%model \sep
%Copula \sep
%Time series

\end{abstract}
\maketitle

%\title{Dynamic models using score copula innovations}
%\author{Landan Zhang, Robert Kohn, Michael Pitt}
%\begin{abstract}

%\end{abstract}
%\maketitle

%\title{GAS-copula model}

%\author[KCLMath]{Landan Zhang}
%%\ead{landan.zhang@kcl.ac.uk}
%\author[KCLMath]{Michael K. Pitt \corref{cor1}}
%\ead{michael.pitt@kcl.ac.uk}
%\cortext[cor1]{Corresponding author.}
%%\fntext[fn1]{Present address: MRC Toxicology Unit, University of Cambridge, Leicester LE1 7HB, United Kingdom \& Email: zh320@mrc-tox.cam.ac.uk}
%\address[KCLMath]{ Department of Mathematics, King’s College London, London WC2R 2LS, United Kingdom}
%\begin{abstract}
%\end{abstract}
%\begin{keyword}
%GAS-copula model \sep
%GARCH model \sep
%Copula \sep
%Time series
%\end{keyword}
%\end{frontmatter}

%\\At this address
%\and Robert Kohn\\At this Address
%\and Michael Pitt\\At this Address}

%\author{Landan Zhang\\At this address}
%\author{Robert Kohn\\At this Address}
%\author{Michael Pitt\\At this Address}

% \linenumbers

%% main text
\clearpage

\section{Introduction}

\label{sec:1} This paper introduces a new class of dynamic models which are
driven by innovations of copula form. In particular, the models are imposed to
be strictly stationary. The observed time series will be denoted as $y_{t}$
for $t=1,..,T$ whilst the time varying parameter is denoted as $\alpha_{t}$. A
useful distinction, following \cite{cox1981statistical} and
\cite{shephard1996statistical}, is to consider time series as being of two
types: observation-driven and parameter-driven models. The first type of model
allows $\alpha_{t}$ to be a function of lagged values of $y_{t}$. For example
the autoregressive conditional heteroscedasticity (ARCH) models introduced by
\cite{engle1982autoregressive} are observation-driven. The advantage of these
models is that the likelihood can be explicitly written down as the
forecasting density for the observations is available. For parameter driven
models a stochastic term is involved in the evolution process for $\alpha_{t}%
$. The stochastic volatility (SV) model (see \cite{shephard1996statistical})
is an example of a parameter driven model. The properties of these latent
variable models are straightforward as the conditions for strict stationarity
are easily verified. The marginal distribution for $y_{t}$ when the model is
stationary is also easily established. However, the likelihood is intractable
and one-step and multi-step prediction densities are usually unavailable.

The new class of models introduced in this article involves the construction
of an observation-driven model. In particular, the models provide the
innovations of the parameter process and use the Generalised Autoregressive
Score (GAS), see \cite{creal2013generalized}, as guidance for the appropriate
copula function. We call this new model class\ innovation GAS\ copula (iGASC)
models. The idea is to take the distribution of the generalized score function
as the variable of interest within a probability transformation. In the case
of multivariate time varying parameters $\alpha_{t}$, the vector score
function may be transformed into a multivariate distribution with uniform
marginals prior to being transformed into a multivariate distribution with
specified marginals. A multivariate distribution with uniform marginals is
known as a copula with a textbook treatment provided by
\cite{joe1997multivariate} and econometric applications given by
\cite{patton2009copula}.

The models we introduce have many of the advantages of parameter-driven
processes whilst retaining the computational convenience of observation driven
models. Firstly, the likelihood is explicitly computed as the one step ahead
forecasts are available. Secondly, Section~\ref{sec:properties} shows that is
straightforward to obtain the marginal distribution of the observations,
whereas this is not the case for most data driven models such as GARCH.
Thirdly, the expression of the parameter process $\alpha_{t}$ as a Gaussian
ARMA($p$,$q$) model is immediate with the constraints for strict stationarity
usually imposed. Fourthly, as a consequence the marginal distribution for any
collection of the initial sequence $(\alpha_{1},...,\alpha_{p})$ is easily
derived and can be used for the initialisation for maximum likelihood (ML)
estimation. Finally, Section~\ref{sec:properties} shows that one-step and
multi-period forecasts are straight-forward.

The structure of this paper is as follows. The univariate approach is
described in Section \ref{sec:uni} and illustrated for a volatility model in
Section \ref{sec:vol}. The marginal and temporal properties of the model is
discussed in Section \ref{sec:properties}. In Section \ref{sec:forecasting},
forecasting multiple periods ahead is considered together with an ARMA
representation for the state transition. The extension to volatility models
for which the standardised return are t-distributed in provided in
\ref{sec:vol_t}. The simple approach for duration models, where the
observations follow a conditionally Exponential distribution, is given in
Section \ref{sec:dur}. The generalization of the duration models to
conditional Weibull distributions is explored in Section \ref{sec:dur_weib}.
In Section \ref{sec:results_uni}, the new iGASC models for volatility are
applied to Nikkei 225 returns at both the daily and weekly frequency. The
multivariate extension of the models is considered in Section
\ref{sec:multivariate}. The multivariate volatility model is described in
Section \ref{sec:mult_SV} and applied to three stock indices in Section
\ref{sec:mult_results}. Finally, in Section \ref{sec:discussion} we conclude.

\section{Univariate models\label{sec:uni}}

This Section considers univariate innovation GAS\ copula (iGASC) models. Let
$y_{t}$ denote a dependent variate with a time-varying parameter $\alpha_{t}$.
Then the simple data generating process (DGP) may be written as
\begin{align}
y_{t}  &  \sim f(y_{t}|\alpha_{t};\nu)\label{eq:uni_model}\\
\alpha_{t+1}  &  =\mu+\phi\alpha_{t}+\psi\eta_{t}.\nonumber
\end{align}
This model may be generalised to be multivariate and to have various lags of
$\alpha_{t}$ and $\eta_{t}$ on the right hand side for the parameter
generating equation. Sections \ref{sec:vol} and \ref{sec:dur} discuss the
crucial issue of constructing the innovation term $\eta_{t}$, which is a
function of current and past data. The four parameters of (\ref{eq:uni_model})
will be represented as $\theta:=(\mu,\phi,\psi,\nu)^{\prime}$.

The innovation in the iGASC model is defined by%
\begin{equation}
\eta_{t}=\Phi^{-1}\left[  F_{g}\left\{  g(y_{t};\alpha_{t},\theta)\mid
\alpha_{t};\theta\right\}  \right]  ; \label{eq:cop_update}%
\end{equation}
$\Phi(\bullet)$ is a standard normal cdf; the univariate function $g$ is
discussed below; $F_{g}$ is the cdf of the random variable $g(Y_{t};\alpha
_{t},\theta)$, where $Y_{t}\sim f(y_{t}|\alpha_{t};\theta)$.

If $y_{t}\sim f(y_{t}|\alpha_{t};\theta)$, so that the DGP of
(\ref{eq:uni_model}) is correct, then $F_{g}\left\{  g(y_{t};\alpha_{t}%
,\theta)\mid\alpha_{t};\theta\right\}  $ are independent and identically
uniformly distributed on $[0,1]$ by the probability integral transform
\citep[see, e.g.][]{angus1994probability}. Hence the innovations $\eta_{t}$
will be standard normal and independent.

Even though the choice of $g$ in (\ref{eq:cop_update}) can be quite general in
principle, its choice in practice depends on two issues. The first is to
choose a for which $F_{g}$ is straightforward to evaluate. The second is that
the function $g$ makes sense in terms of the relationship between the
observations and the innovations. In general the choice of $g$ is taken to be
the score function%
\begin{equation}
g(y_{t};\alpha_{t},\theta)=\frac{\partial\ln f(y_{t}\mid\alpha_{t};\theta
)}{\partial\alpha_{t}}, \label{eq:score}%
\end{equation}
This choice is motivated by the associated GAS literature, see
\citet{creal2011dynamic} and \cite{creal2013generalized}. Sections
\ref{sec:vol} and \ref{sec:dur} show that this choice typically satisfies both criteria.

\subsection{Volatility model: conditionally Gaussian\label{sec:vol}}

Consider the conditional Gaussian model for the return $y_{t}=\sigma
_{t}\varepsilon_{t}$, where $\varepsilon_{t}\overset{iid}{\sim}\mathcal{N}%
(0,1)$. Let $\alpha_{t}=\ln\sigma_{t}^{2}$ be the time-varying log-volatility.
Then
\begin{equation}
y_{t}\mid\alpha_{t}\sim\mathcal{N}(0,e^{\alpha_{t}}),\quad\text{and}%
\quad\alpha_{t+1}=\mu+\phi\alpha_{t}+\psi\eta_{t}. \label{eq:vol}%
\end{equation}
This is a special case of the model of (\ref{eq:uni_model}), where the term
$\alpha_{t}$ is the log-variance. The model is similar in appearance to both
the stochastic volatility model; see \cite{shephard1996statistical}, as well
as the EGARCH\ model of \cite{nelson1991conditional}. For the EGARCH\ model,
the innovations are structured to be a Martingale Difference (MD) sequence.

The innovations $\eta_{t}$ are defined in (\ref{eq:cop_update}), where the
function $g$ is taken to be the score function of (\ref{eq:score}). For this
model,
\begin{align*}
\ln f(y_{t}  &  \mid\alpha_{t};\theta)=-\frac{1}{2}\ln({2\pi})-\frac{1}%
{2}y_{t}^{2}e^{-\alpha_{t}}-\frac{1}{2}\alpha_{t}.\\
g(y_{t};\alpha_{t},\theta)  &  =\frac{\partial\ln f(y_{t}\mid\alpha_{t}%
;\theta)}{\partial\alpha_{t}}\\
&  =\frac{1}{2}y_{t}^{2}e^{-\alpha_{t}}-\frac{1}{2}=\frac{1}{2}(\varepsilon
_{t}^{2}-1).
\end{align*}
The distribution function $F_{g}\left\{  g(y_{t};\alpha_{t},\theta)\mid
\alpha_{t};\theta\right\}  $ is therefore%
\[
F_{g}\left\{  g(y_{t};\alpha_{t},\theta)\mid\alpha_{t};\theta\right\}
=\Pr\left(  Z^{2}\leq\varepsilon_{t}^{2}\right)  =F_{\chi_{1}^{2}}%
(\varepsilon_{t}^{2}),
\]
where $Z$ is standard normal and $F_{\chi_{1}^{2}}$ is the distribution
function for the chi-square distribution with one degree of freedom.

\begin{figure}[th]
\centering
\includegraphics[width=0.8\textwidth]{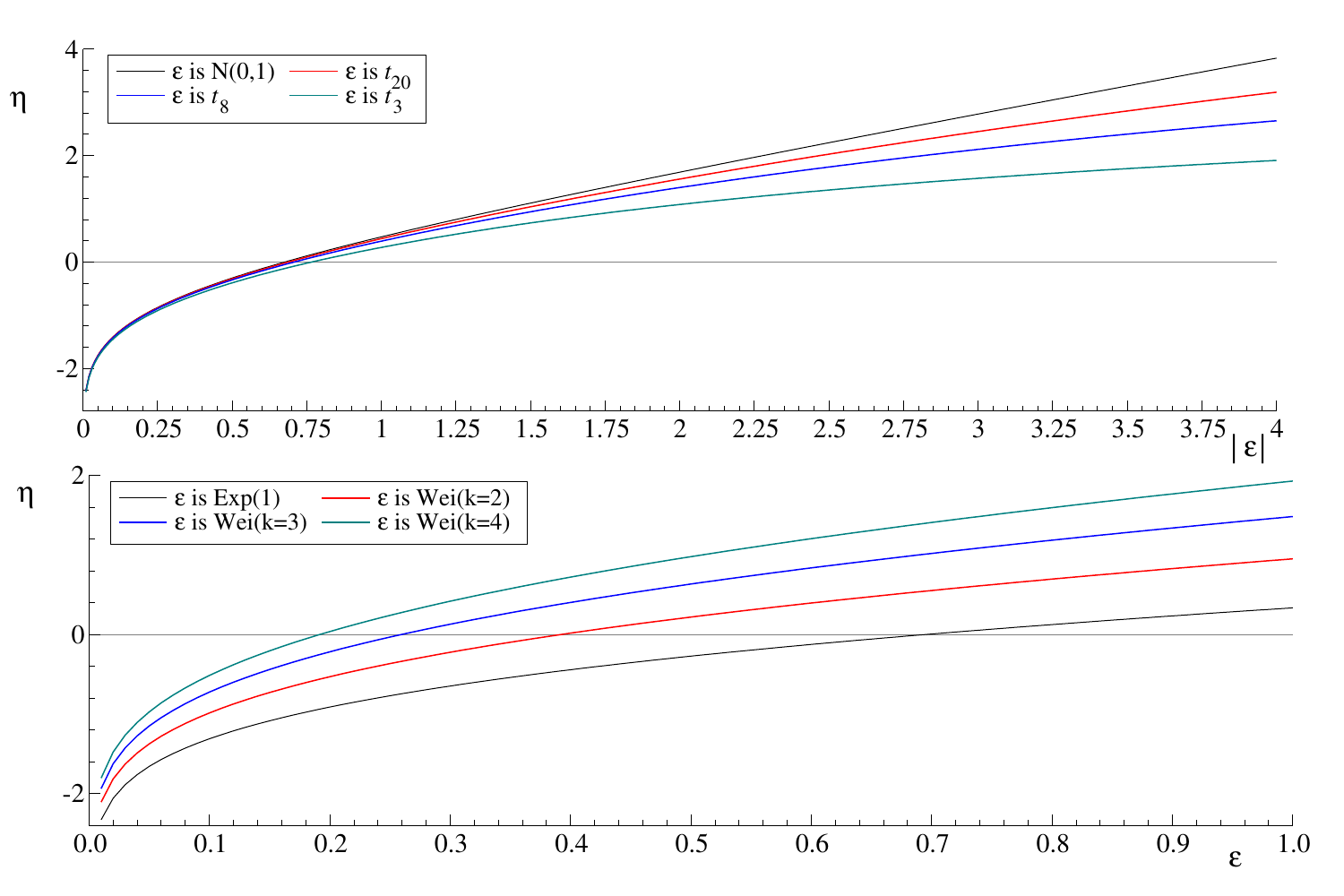}\caption{\textsl{The
relationship between }$\eta_{t}$\textsl{ and }$\varepsilon_{t}$. \textsl{TOP:
}$\eta_{t}$ \textsl{against }$\left\vert \varepsilon_{t}\right\vert $\textsl{
for the volatility model where }$\varepsilon_{t}$\textsl{ is standard Gaussian
and }$t_{\nu}$\textsl{. BOTTOM: }$\eta_{t}$\textsl{ against }$\varepsilon
$\textsl{ where }$\varepsilon_{t}$\textsl{ is standard Exponential, Weibull
(for }$k=2$\textsl{ to 4) for the duration model. }}%
\label{fig:epseta}%
\end{figure}

In this case the score approach, suggested by the GAS models, for the function
$g$ appears appropriate. The innovation term, defined in (\ref{eq:cop_update}%
),
\begin{equation}
\eta_{t}=\Phi^{-1}\left(  F_{\chi_{1}^{2}}(\varepsilon_{t}^{2})\right)
\label{eq:vol_innov}%
\end{equation}
is easily calculated. and $\eta_{t}$ is a strictly monotonically increasing
function of the magnitude of standardised returns $\left\vert \varepsilon
_{t}\right\vert $. The top panel of Figure \ref{fig:epseta} plots the
relationship between the two; this is very similar to the form of EGARCH
processes, \cite{nelson1991conditional} when an asymmetric term is absent.
This is consistent with the belief that larger absolute standardised returns
$\left\vert \varepsilon_{t}\right\vert $ lead to larger variance at the
following time point.

The likelihood is easily calculated by using the prediction decomposition as
the forecast distribution is readily available by construction. For a time
series $y_{1:T}=(y_{1},...,y_{T})^{\prime}$ the forecast density $f(y_{t}$ $|$
$y_{1:t-1};\theta)=f(y_{t}\mid\alpha_{t};\theta)$, as $\alpha_{t}$ is
constructed from the previous values $y_{1:t-1}$. The log-likelihood is
calculated via the prediction decomposition as%

\begin{align}
\ell(\theta)  &  =\sum_{t=1}^{T}\log f(y_{t}|y_{1:t-1};\theta
)\label{eq:loglik_vol}\\
&  =-\frac{1}{2}\sum_{t=1}^{T}y_{t}^{2}e^{-\alpha_{t}}-\frac{1}{2}\sum
_{t=1}^{T}\alpha_{t}-\frac{T}{2}\ln({2\pi}),\nonumber
\end{align}
where the log-variance $\alpha_{t}$ is given by the evolution of
(\ref{eq:vol}) and the calculation of $\eta_{t}$ through (\ref{eq:vol_innov}).

In practice, we add a very small positive constant offset term to
$\varepsilon_{t}^{2}$ of $10^{-4}$ when returns are expressed as percentages.
This avoid any numerical instability in the calculation of (\ref{eq:vol_innov}%
) for returns which are exactly zero. This is exactly the tiny correction term
used to prevent instability when calculating $\log y_{t}^{2}$, see for example
\cite{harvey1994multivariate}.

\subsubsection{Properties of iGASc models\label{sec:properties}}

The iGASc model has the advantages of the observation driven class of time
series models. Specifically, the forecasting density can be evaluated enabling
maximum likelihood as seen in (\ref{eq:loglik_vol}). The iGASc also shares
many attractive properties of the parameter driven class of models. The
marginal distribution of both the states $\alpha_{t}$ and the observations
$y_{t}$ may be easily derived. Consider the model of (\ref{eq:vol}) then the
process for $\alpha_{t}$ is simply a Gaussian autoregression and provided that
$\left\vert \phi\right\vert <1$,
\begin{align}
\alpha_{t}  &  \sim\mathcal{N}(\mu_{\alpha},\sigma_{\alpha}^{2}%
),\label{eq:marg_state}\\
\mu_{\alpha}  &  =\frac{\mu}{1-\phi},\text{ }\sigma_{\alpha}^{2}=\frac
{\psi^{2}}{1-\phi^{2}}.\nonumber
\end{align}
This in turn means that the marginal density for the observations can be
computed as a simply univariate mixture
\[
f(y_{t})=%
%TCIMACRO{\dint }%
%BeginExpansion
{\displaystyle\int}
%EndExpansion
\mathcal{N}(y_{t}|0,e^{\alpha_{t}})\mathcal{N}(\alpha_{t}|\mu_{\alpha}%
,\sigma_{\alpha}^{2})d\alpha_{t}.
\]
This is very similar to the properties of the parameter driven stochastic
volatility (SV) model; see \cite{shephard1996statistical}. This means the
marginal properties of the observation process are readily available. For
example, the kurtosis of the returns may be calculated as
\[
K_{y}=\frac{E[y_{t}^{4}]}{E[y_{t}^{2}]^{2}}=3e^{\sigma_{\alpha}^{2}}.
\]
The temporal properties of the model are also similar to SV models. In
particular the autocorrelation function may be derived by considering the
unobserved components representation of (\ref{eq:vol}) as
\[
\log(y_{t}^{2})=\alpha_{t}+\log(\varepsilon_{t}^{2}),
\]
where, see \cite{shephard1996statistical}, the error $\log(\varepsilon_{t}%
^{2})$ is a log chi-squared variable with variance $4.93$. We note that
$\varepsilon_{t}$ is independent of $\alpha_{t}$ as the latter is constructed
from previous information $y_{1:t-1}=(y_{1},...,y_{t-1})^{\prime}$. We
therefore obtain that the autocorrelation for the series $\log(y_{t}^{2})$ at
lag $\tau$ is
\[
\rho(\tau)=\frac{\phi^{\tau}\sigma_{\alpha}^{2}}{\sigma_{\alpha}^{2}+4.93}.
\]
Whilst the properties of the iGASc models have been illustrated for the simple
volatility model of (\ref{eq:vol}), these properties apply more generally, for
example to the duration model of Section \ref{sec:dur}.

\subsubsection{Forecasting and ARMA representation\label{sec:forecasting}}

As discussed in Section \ref{sec:vol}, the state $\alpha_{t}$ is formed from
past observations $y_{1:t-1}=(y_{1},...,y_{t-1})^{\prime}$ and the model
parameters $\theta$. Forecasting one step ahead is therefore explicit as
$f(y_{t}$ $|$ $y_{1:t-1};\theta)=f(y_{t}\mid\alpha_{t};\theta)$. To illustrate
forecasting several steps ahead, the model of (\ref{eq:uni_model}) is
considered. In this case,
\begin{equation}
\alpha_{t+h}\mid y_{1:t};\theta\sim\mathcal{N}\left(  \phi^{h-1}\alpha
_{t}+\{1-\phi^{h-1}\}\mu_{\alpha},\{1-\phi^{2(h-1)}\}\sigma_{\alpha}%
^{2}\right)  , \label{eq:forecast_state}%
\end{equation}
where the quantities $\mu_{\alpha},$ $\sigma_{\alpha}^{2}$ are given in
(\ref{eq:marg_state}). The forecast of $y_{t+h}$ is therefore a simple
univariate mixture%
\[
f(y_{t+h}\mid y_{1:t};\theta)=%
%TCIMACRO{\dint }%
%BeginExpansion
{\displaystyle\int}
%EndExpansion
f(y_{t+h}\mid\alpha_{t+h};\theta)f(\alpha_{t+h}\mid y_{1:t};\theta
)d\alpha_{t+h},
\]
where the state forecast density is given by (\ref{eq:forecast_state}). The
calculation of moments of $y_{t+h}$ under $f(y_{t+h}\mid y_{1:t};\theta)$ is
typically straight-forward.

The model is readily extended to an ARMA($p$,$q$) process in $\alpha_{t}$ by
writing
\begin{equation}
\alpha_{t+1}=\mu+\sum_{i=1}^{p}\phi_{i}\alpha_{t-1+1}+\sum_{j=1}^{q}\psi
_{j}\eta_{t-j}+\psi_{0}\eta_{t}; \label{eq:vol_ARMA}%
\end{equation}
where the roots of the lag polynomial $\phi(z)=1-\phi_{1}z-\phi_{2}%
z^{2}-...-\phi_{p}z^{p}$ need to lie outside the unit circle for the process
to be stationary. The innovation terms $\eta_{t}$ and $\eta_{t-j}$ are again
constructed as in \ (\ref{eq:cop_update}), and for the volatility model the
innovations are calculated according to (\ref{eq:vol_innov}). Strict
stationarity coincides with weak stationarity when the innovations $\eta_{t}$
are Gaussian.

\subsection{Volatility models: heavy tailed extensions\label{sec:vol_t}}

A natural extension of the volatility model of Section \ref{sec:vol} is to
assume that the standardised returns $\varepsilon_{t}$ have a heavier tailed
distribution than a standard Gaussian variate. For example, $y_{t}=\sigma
_{t}\varepsilon_{t}$, where $\varepsilon_{t}$ is now an independent and
identically (IID) $t$-distributed variate with $\nu$ degrees of freedom and
unit variance. The density of $\varepsilon_{t}$ is given as%
\[
f_{\varepsilon}(\varepsilon_{t})=c\left(  1+\frac{\varepsilon_{t}^{2}}{\nu
-2}\right)  ^{-\frac{\nu+1}{2}},\text{ }c=\frac{\Gamma\left(  \frac{\nu+1}%
{2}\right)  }{\Gamma\left(  \frac{\nu}{2}\right)  \sqrt{(\nu-2)\pi}}.
\]
Similarly to (\ref{eq:vol}) where $\alpha_{t}=\ln\sigma_{t}^{2}$,%
\[
f(y_{t}\mid\alpha_{t};\theta)=e^{-\frac{\alpha_{t}}{2}}f_{\varepsilon}\left(
y_{t}e^{-\frac{\alpha_{t}}{2}}\right)  .
\]
The score function is%
\[
g(y_{t};\alpha_{t},\theta)=\frac{\partial\ln f(y_{t}\mid\alpha_{t};\theta
)}{\partial\alpha_{t}}=-\frac{1}{2}+\frac{(\nu+1)}{2}\frac{1}{(\nu
-2)\varepsilon_{t}^{-2}+1}.
\]
The corresponding distribution function is%
\[
F_{g}\left\{  g(y_{t};\alpha_{t},\theta)\mid\alpha_{t};\theta\right\}
=\Pr\left(  T_{\nu}^{2}\leq\frac{\nu}{\nu-2}\varepsilon_{t}^{2}\right)  ,
\]
where $T_{\nu}$ is a random variable following a $t$-distribution with $\nu$
degrees of freedom. That is,%
\[
F_{g}\left\{  g(y_{t};\alpha_{t},\theta)\mid\alpha_{t};\theta\right\}
=F_{\mathcal{F}_{1,\nu}}\left(  \frac{\nu}{\nu-2}\varepsilon_{t}^{2}\right)
,\text{ }\eta_{t}=\Phi^{-1}\left\{  F_{\mathcal{F}_{1,\nu}}\left(  \frac{\nu
}{\nu-2}\varepsilon_{t}^{2}\right)  \right\}  .
\]
where $F_{\mathcal{F}_{1,\nu}}$ is the distribution function for the
$F$-distribution with parameters $1$ and $\nu$ degrees of freedom. Similarly
to the volatility model with Gaussian innovations $\varepsilon_{t}$, the
innovation term $\eta_{t}$ is again monotonically increasing as a function of
the standardised return $\left\vert \varepsilon_{t}\right\vert $. The top
panel of Figure \ref{fig:epseta} plots the relationship between the two, for
various values of $\nu$.

The forecasting density and likelihood function can be calculated
straightforwardly similarly to (\ref{eq:loglik_vol}). The extension to a
Gaussian ARMA($p,q$) model of (\ref{eq:vol_ARMA}) is immediate as the
innovations $\eta_{t}$ are again standard Gaussian.

\subsection{Duration models\label{sec:dur}}

The iGASC model can be used to model duration data. The conditional model will
be used to model the time between successive trades $y_{i}=\tau_{i}-\tau
_{i-1}$, with $\tau_{i}$ representing the time of the $i$-th trade for
$i=1,2,...,N$. Here, $\tau_{0}$ and $\tau_{N}$ represent the time of the first
and last trade respectively. For this model, $y_{i}=\lambda_{i}\varepsilon
_{i}$ where, following the approach of \cite{engle1998autoregressive},
$\varepsilon_{i}\overset{iid}{\sim}Exp(1)$, a standard Exponential
distribution. Transforming to the real line $\alpha_{i}=\ln\lambda_{i}$, the
conditional density is
\[
p(y_{i}\mid\alpha_{i};\theta)=\frac{1}{\lambda_{i}}e^{-y_{i}/\lambda_{i}}%
=\exp\left(  -\alpha_{i}-y_{i}e^{-\alpha_{i}}\right)  .
\]
The parameter evolution is%
\[
\alpha_{i+1}=\mu+\phi\alpha_{i}+\psi\eta_{i}.
\]
We now construct the innovation term $\eta_{i}$. The conditional score is
\begin{align*}
g(y_{i};\alpha_{i},\theta)  &  =\frac{\partial\ln p(y_{i}\mid\alpha_{i}%
;\theta)}{\partial\alpha_{i}}=-1+y_{i}e^{-\alpha_{i}}\\
&  =-1+\varepsilon_{i}.
\end{align*}
Therefore, the distribution function is%
\[
F_{g}\left\{  g(y_{i};\alpha_{i},\theta)\mid\alpha_{i};\theta\right\}
=F_{\varepsilon}(\varepsilon_{i})=1-e^{-\varepsilon_{i}},
\]
where $F_{\varepsilon}$ is the distribution function for the standard
exponential distribution. Therefore, the innovation term is
\[
\text{ }\eta_{i}=\Phi^{-1}\left\{  F_{\varepsilon}(\varepsilon_{i})\right\}
=\Phi^{-1}\left(  1-e^{-\varepsilon_{i}}\right)  .
\]
Clearly, the innovation $\eta_{i}$ is a monotonically increasing function of
the standardised duration $\varepsilon_{i}$. The bottom panel of Figure
\ref{fig:epseta} plots relationship between the two variables, for various
values of $\nu$. This model therefore has a qualitative similarity to that of
\cite{engle1998autoregressive} for which the durations themselves are used to
drive the parameter process. The advantages of the iGASC model are again that
the marginal distribution is simply calculated and the forecasts are of closed
distributional form even many steps ahead.

\subsubsection{Duration models: conditional Weibull
variates\label{sec:dur_weib}}

The extension to a conditional Weibull, a generalisation of the Exponential
distribution. In this case $\varepsilon_{i}\overset{iid}{\sim}Wei(\beta,k)$
where $\beta=1/\Gamma(1+1/k)$ to ensure $E[\varepsilon_{i}]=1$. This
standardised variate is therefore indexed by a single parameter $k$. The
density of $\varepsilon_{i}$ is
\[
f_{\varepsilon}(\varepsilon_{i})=\frac{k}{\beta}\left(  \frac{\varepsilon_{i}%
}{\beta}\right)  ^{k-1}\exp\left\{  -\left(  \frac{\varepsilon_{i}}{\beta
}\right)  ^{k}\right\}  .
\]
When $k=1$, this simply reduces to the conditional Exponential model. As
before $y_{i}=\lambda_{i}\varepsilon_{i}$ following the approach of
\cite{engle1998autoregressive}. Again, taking $\alpha_{i}=\ln\lambda_{i}$,
\begin{align*}
p(y_{i}  &  \mid\alpha_{i};\theta)=e^{-\alpha_{i}}f_{\varepsilon}\left(
y_{i}e^{-\alpha_{i}}\right) \\
\log p(y_{i}  &  \mid\alpha_{i};\theta)=const-\alpha_{i}+(k-1)\log
(y_{i}e^{-\alpha_{i}})-\left(  \frac{y_{i}e^{-\alpha_{i}}}{\beta}\right)
^{k}\\
&  =const-k\alpha_{i}-\left(  \frac{y_{i}}{\beta}e^{-\alpha_{i}}\right)
^{k}=const-k\alpha_{i}-\left(  \frac{\varepsilon_{i}}{\beta}\right)  ^{k}\\
g(y_{i};\alpha_{i},\theta)  &  =\frac{\partial\ln p(y_{i}\mid\alpha_{i}%
;\theta)}{\partial\alpha_{i}}=-k+k\left(  \frac{y_{i}}{\beta}e^{-\alpha_{i}%
}\right)  ^{k}=k\frac{\varepsilon_{i}^{k}}{\beta^{k}}-k
\end{align*}
Now consider the distribution function $F_{g}\{g(y_{i})\}$ of $g(y_{i})$ under
the conditional density $p(y_{i}\mid\alpha_{i};\theta),$ where we denote
$\epsilon$ as a $Wei(\beta,k)$ random variable,
\begin{align*}
F_{g}\left\{  g(y_{i};\alpha_{i},\theta)\mid\alpha_{i};\theta\right\}   &
=\Pr(\epsilon^{k}\leq\varepsilon_{i}^{k})=\Pr(\epsilon\leq\varepsilon
_{i})=1-\exp\left\{  -\left(  \frac{\varepsilon_{i}}{\beta}\right)
^{k}\right\}  ,\\
\eta_{i}  &  =\Phi^{-1}\left\{  F_{\varepsilon}(\varepsilon_{i})\right\}
=\Phi^{-1}\left(  1-\exp\left\{  -\left(  \frac{\varepsilon_{i}}{\beta
}\right)  ^{k}\right\}  \right)  .
\end{align*}
The Gaussian innovation $\eta_{i}$ as a function of the standardised duration
$\varepsilon_{i}$ is show in Figure \ref{fig:epseta} for various values of
$k$. The iGASC model Weibull duration model reduces to the conditionally
Exponential model when $k=1$ and so $\beta=1/\Gamma(1+1/k)=1.$

The duration model the extension to a Gaussian ARMA($p,q$) model of
(\ref{eq:vol_ARMA}) is straight-forward as the innovations $\eta_{t}$ are
again standard Gaussian. The likelihood is readily available by using the
prediction decomposition.

\section{\bigskip Results: univariate model\label{sec:results_uni}}

\subsection{Synthetic data}

\begin{table}[th]
\begin{center}%
\begin{tabular}
[c]{c|cccc}\hline
Parameter & $\mu$ & $\phi$ & $\psi$ & $\nu$\\\hline
True value & 0.3 & 0.2 & 0.7 & 10\\\hline
$T=1000$ &  &  &  & \\
Mean & 0.29348 & 0.20431 & 0.69805 & 11.70043\\
Variance & 0.00587 & 0.00678 & 0.00286 & 24.01240\\
Bias & 0.00651 & 0.00431 & 0.00194 & 1.70043\\
MSE & 0.00586 & 0.00673 & 0.00283 & 26.66371\\\hline
$T=5000$ &  &  &  & \\
Mean & 0.30068 & 0.19969 & 0.70081 & 10.23158\\
Variance & 0.00105 & 0.00116 & 0.00041 & 1.57374\\
Bias & 0.00068 & 0.00030 & 0.00081 & 0.23158\\
MSE & 0.00104 & 0.00115 & 0.00041 & 1.61164\\\hline
$T=10000$ &  &  &  & \\
Mean & 0.29759 & 0.20179 & 0.69968 & 10.24054\\
Variance & 0.00057 & 0.00055 & 0.00033 & 1.08449\\
Bias & 0.00240 & 0.00179 & 0.00031 & 0.24054\\
MSE & 0.00057 & 0.00055 & 0.00032 & 1.13150\\\hline
$T=20000$ &  &  &  & \\
Mean & 0.29764 & 0.20149 & 0.70072 & 10.03524\\
Variance & 0.00017 & 0.00026 & 0.00011 & 0.38072\\
Bias & 0.00235 & 0.00149 & 0.00072 & 0.03524\\
MSE & 0.00017 & 0.00026 & 0.00011 & 0.37815\\\hline
\end{tabular}
\end{center}
\caption{\textsl{Results over 200 replications of the univariate iGASC-t
volatility model of Section \ref{sec:vol_t}. The mean, variance, bias and MSE
of the resulting estimates are shown. Different lengths }$T$ \textsl{for the
time series are used.}}%
\label{tab:montecarlo}%
\end{table}

Simulation studies are carried out for the univariate iGASC-t volatility
models of Section \ref{sec:vol_t}. The parameters are $\mu=0.3$, $\phi=0.2$,
$\psi=0.7$ and $\nu=10$. The replications are over $200$ data sets of length
$T=1000,5000,10000$ and $20000$, respectively. Maximum likelihood is used to
estimate the parameters of interest and the mean, variance, bias and mean
squared errors (MSE) of the resulting estimates are summarised in
Table~\ref{tab:montecarlo}. As shown in Table~\ref{tab:montecarlo}, the method
produces accurate estimates for the parameter of interests. With the increase
of the length of the time period $T$, the variation of the resulting estimates
significantly decreases.\begin{table}[tb]
\centering
\begin{tabular}
[c]{cccc}\hline
iGASC & GARCH(1,1) & iGASC & GARCH(1,1)\\
Gaussian & Gaussian & Student t & Student t\\\hline
$\mu$ & $\omega$ & $\mu$ & $\omega$\\
0.03568 & 0.05761 & 0.01651 & 0.04235\\
(0.02334, 0.04802) & (0.02939, 0.08583) & (0.00582, 0.02720) & (0.01301,
0.07169)\\
$\phi$ & $\alpha$ & $\phi$ & $\alpha$\\
0.95231 & 0.13411 & 0.96651 & 0.11838\\
(0.93347, 0.97115) & (0.10221, 0.16601) & (0.94794, 0.98508) & (0.07844
0.15832)\\
$\psi$ & $\beta$ & $\psi$ & $\beta$\\
0.16096 & 0.83982 & 0.14169 & 0.86439\\
(0.13140, 0.19052) & (0.80158, 0.87806) & (0.10322, 0.18016) & (0.81875,
0.91003)\\
&  & $\nu$ & $\nu$\\
&  & 6.29119 & 6.27660\\
&  & (4.75189, 7.83049) & (4.72964, 7.82356)\\\hline
\end{tabular}
\caption{\textsl{Parameter estimates and 95\% confidence intervals for iGASC
and GARCH(1,1) models applied to daily Nikkei returns. The distribution of the
standardised return }$\varepsilon_{t}$\textsl{ is considered as Gaussian or
t--distributed.}}%
\label{tab:param_nikkei}%
\end{table}

\subsection{Return data: Nikkei index\label{sec:ret_data}}

\begin{figure}[th]
\centering
\includegraphics[width=0.9\textwidth]{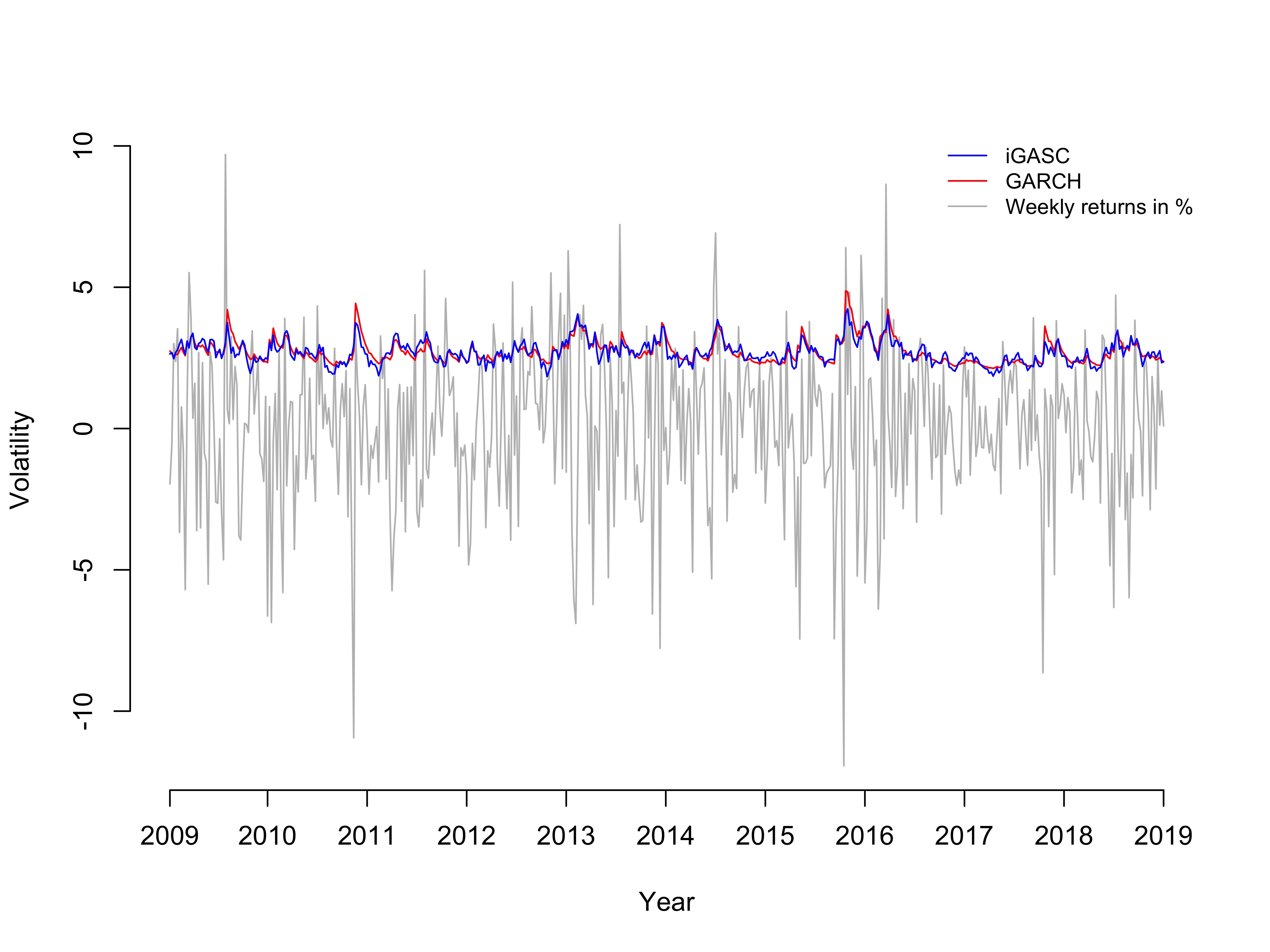} \caption{\textsl{Results
for the 520 weekly continuously compounded percentage returns for the Nikkei.
The standard deviations }$\sigma_{t}$ \textsl{for the GARCH(1,1)-t model and
}$\sigma_{t}=\exp(\alpha_{t}/2)$\textsl{ for the iGASC-t model, of Sec
\ref{sec:vol_t}, are displayed with the return.}}%
\label{fig:vol_t}%
\end{figure}

\begin{table}[ptb]
\centering%
\begin{tabular}
[c]{cccc}\hline
iGASC & GARCH(1,1) & iGASC & GARCH(1,1)\\
Gaussian & Gaussian & Student t & Student t\\\hline
$\mu$ & $\omega$ & $\mu$ & $\omega$\\
0.49686 & 1.78208 & 0.33798 & 1.04019\\
(0.05370, 0.94002) & (-0.74113, 4.30529) & (-0.11183 0.78779) & (-0.63373,
2.71411)\\
$\phi$ & $\alpha$ & $\phi$ & $\alpha$\\
0.74993 & 0.15256 & 0.82583 & 0.10748\\
(0.52614, 0.97372) & (0.01837, 0.28675) & (0.59707, 1.05459) & (-0.01055,
0.22551)\\
$\psi$ & $\beta$ & $\psi$ & $\beta$\\
0.21434 & 0.61593 & 0.18103 & 0.75617\\
(0.10272, 0.32596) & (0.19244, 1.03942) & (0.05475 0.30731) & (0.45099,
1.06135)\\
&  & $\nu$ & $\nu$\\
&  & 9.98325 & 8.62882\\
&  & (2.52799 17.43851) & (2.93902, 14.31862)\\\hline
\end{tabular}
\caption{\textsl{Parameter estimates and 95\% confidence intervals for iGASC
and GARCH(1,1) models applied to weekly Nikkei returns. The distribution of
the standardised return }$\varepsilon_{t}$\textsl{ is considered as Gaussian
or t--distributed.}}%
\label{tab:param_nikkei_week}%
\end{table}The iGASC volatility models of Section \ref{sec:vol} are applied to
the daily closing prices for the Nikkei 225.\ The two different conditional
densities for the returns, standard Gaussian and standard t, are considered.
These are compared to the GARCH(1,1) and GARCH(1,1)-t models, see
\cite{bollerslev1987conditionally}. The returns are continuously compounded
and calculated as $y_{t}=100(\log S_{t}-\log S_{t-1}$), where $S_{t}$
represents the closing price of the stock index. The data is from 2009-04-30
to 2019-04-30 leading to 2476 daily and 520 weekly percentage returns. The
parameter estimates for the daily returns are provided in Table
\ref{tab:param_nikkei}. It is clear that the persistence for the GARCH class,
given by $(\alpha+\beta)$, is a little higher than the persistence, given by
$\phi$, for the two iGASC models. The estimates for the degrees of freedom
parameter $\nu$, for the conditional distribution of returns, are very similar
for both the GARCH and iGASC model. The corresponding estimates for the 520
weekly returns are provided in Table \ref{tab:param_nikkei_week}. It is again
apparent that the iGASC models result in less persistence when compared to the
corresponding GARCH models. However, the degrees of freedom parameter $\nu$ is
estimated to be higher for the iGASC-t model than for the GARCH-t model,
although the confidence intervals are wide in both cases. The standard
deviations $\sigma_{t}$ for the GARCH(1,1)-t model and $\sigma_{t}=\exp
(\alpha_{t}/2)$ for the iGASC-t volatility model are displayed for the weekly
Nikkei returns in Figure \ref{fig:vol_t}. \ \begin{table}[ptb]
\centering
\begin{tabular}
[c]{c|cccc}\hline
Model & iGASC & GARCH & iGASC & GARCH\\
Standardised return & Gaussian & Gaussian & Student t & Student t\\\hline
Daily & -4000.082 & -4002.294 & -3941.135 & -3944.288\\
Weekly & -1249.165 & -1252.573 & -1244.543 & -1246.397\\\hline
\multicolumn{5}{c}{Kolmogorov-Smirnov test (daily series)}\\\hline
Statistic D & 0.047565 & 0.049637 & 0.026795 & 0.030647\\
p-value & 2.725e-05 & 1.005e-05 & 0.05714 & 0.0191\\\hline
\multicolumn{5}{c}{Kolmogorov-Smirnov test (weekly series)}\\\hline
Statistic D & 0.036716 & 0.038501 & 0.041502 & 0.040686\\
p-value & 0.4849 & 0.4239 & 0.3319 & 0.3555\\\hline
\end{tabular}
\caption{\textsl{Results for 2476 daily and 520 weekly percentage returns of
Nikkei 225. The distribution of the standardised return }$\varepsilon_{t}%
$\textsl{ is considered as Gaussian or t--distributed. The maximized
log-likelihoods and the Kolmogorov-Smirnov residual tests are reported for the
four series.}}%
\label{tab:nikkei_loglik}%
\end{table}Diagnostics may be performed for the fit of the models by using
the probability integral transform (PIT) method. The forecast cumulative
distribution function is evaluated as
\[
u_{t}=F(y_{t}\mid y_{1:t-1};\theta)=\Phi\left(  y_{t}e^{-\alpha_{t}/2}\right)
.
\]
Under the assumption that the model and parameters are correct, the statistics
$u_{t}\overset{IID}{\sim}Un(0,1)$ for $t=1,...,T$. A Kolmogorov-Smirnov test
may be applied to test whether the residuals $u_{t}$ are marginally uniform.
The results (p-values) of the tests for the four models under consideration
are summarised in Table \ref{tab:nikkei_loglik} for both the daily and weekly
Nikkei returns. All four models are acceptable under this test at the weekly
frequency. However, only the iGASC-t volatility model is acceptable (at $5\%$
\ size) for daily data. Both the conditionally Gaussian and conditionally
t-distribution models outperform their GARCH\ counterparts in terms of the
log-likelihood at the weekly and daily frequency. These log-likelihood
comparisons are given in Table \ref{tab:nikkei_loglik}.

\section{Multivariate models\label{sec:multivariate}}

\subsection{The copula distribution\label{sec:copula}}

The extension of the model to multivariate states is, in principle,
straightforward. The multivariate version of the probability transform
approach, of (\ref{eq:uni_model}) in Section \ref{sec:uni}, leads to a copula
distribution function. A good review of copula distributions, which are simply
multivariate distributions with univariate marginals is given by, for example,
\cite{joe1997multivariate}.

A copula $C\left(  u_{1},u_{2},\ldots,u_{N}\right)  $ is defined as the joint
cumulative distribution function of variables $\left(  U_{1},U_{2}%
,\ldots,U_{N}\right)  $ where each $0\leq U_{i}\leq1$ for $i=1,...,N$. So
\begin{equation}
C\left(  u_{1},u_{2},\ldots,u_{N}\right)  =\operatorname{Pr}\left(  U_{1}\leq
u_{1},U_{2}\leq u_{2},\ldots,U_{N}\leq u_{N}\right)  . \label{eq:dist_copula}%
\end{equation}
Each marginal distribution is uniformly distributed on the interval $[0,1]$ so
$\operatorname{Pr}\left(  U_{i}\leq u_{i}\right)  =u_{i}$ for $i=1,...,N$.

The use of copulae is widespread as the uniform variables can then be mapped
into any distribution of interest, for example to extreme value distributions
using the probability integral transform. For the purposes of this paper, the
copula will be formed by considering a $N\times1$ continuous random vector
$X$, where the marginal of each component of $X$ are known. In the standard
copula approach\textsl{ }the random vector
\[
\left(  U_{1},U_{2},\ldots,U_{N}\right)  =\left(  F_{1}\left(  X_{1}\right)
,F_{2}\left(  X_{2}\right)  ,\ldots,F_{N}\left(  X_{N}\right)  \right)  ,
\]
where $F_{i}(x)=\operatorname{Pr}(X_{i}\leq x)$. In this case, the copula
function which we shall call $C_{g}\left(  \cdot\right)  $ can be written as
\begin{align*}
C_{g}\left(  u_{1},u_{2},\ldots,u_{d}\right)   &  =\operatorname{Pr}[X_{1}\leq
F_{1}^{-1}\left(  u_{1}\right)  ,X_{2}\leq F_{2}^{-1}\left(  u_{2}\right)
,\ldots,X_{N}\leq F_{N}^{-1}\left(  u_{N}\right)  ]\\
&  =\operatorname{Pr}[F_{1}(X_{1})\leq u_{1},F_{2}\left(  X_{2}\right)  \leq
u_{2},\ldots,F_{N}\left(  X_{N}\right)  \leq u_{N}].
\end{align*}
Perhaps an easier way of considering this copula is to consider generating
random variates from $C_{g}\left(  \cdot\right)  $. The first step would be to
generate the vector $X$ from its distribution function and then simply to
evaluate $U_{i}=F_{i}(X_{i})$ for $i=1,...,N$.

\subsection{Multivariate stochastic volatility models\label{sec:mult_SV}}

We now consider the iGASC approach for the modelling of multivariate
stochastic volatility models. The modelling is similar to the parameter driven
approach taken by \cite{harvey1994multivariate}. The $N\times1$ vector of
continuous compounded returns $y_{t}$ is modelled as jointly Gaussian,
\begin{equation}
y_{it}=e^{\alpha_{it}/2}\varepsilon_{it},\text{ }\varepsilon_{t}%
\sim\mathcal{N}_{N}(0;\Sigma), \label{eq:mult_sv}%
\end{equation}
where $\Sigma$ is a positive definite correlation matrix with $N(N-1)/2$
elements and has $1^{\prime}$s down the diagonal. Equivalently,
\[
y_{t}\mid\alpha_{t};\theta\text{ }\sim\mathcal{N}_{N}(0;D\Sigma D),
\]
where $D$ is a diagonal matrix with elements $D_{ii}=e^{\alpha_{it}/2}$. In a
departure from the parameter driven model, the iGASC model has%
\[
\alpha_{it+1}=\mu_{i}+\phi_{i}\alpha_{it}+\psi_{i}\eta_{it},
\]
where the $\eta_{it}$ innovations terms need to be specified.

The simplest approach is to consider the marginal conditional score function.
That is, we consider
\begin{equation}
g(y_{it};\alpha_{it},\theta)=\frac{\partial\ln f(y_{it}\mid\alpha_{it}%
;\theta)}{\partial\alpha_{it}},\label{eq:marg_score}%
\end{equation}
where $f(y_{it}\mid\alpha_{it},\mathcal{F}_{t-1};\theta)$ represents the
marginal distribution of $y_{it}$ conditional upon $\alpha_{it}$. In this
case,
\[
y_{it}\mid\alpha_{it};\theta\sim\mathcal{N}(0;e^{\alpha_{it}}),
\]
and we consider the distribution of (\ref{eq:marg_score}) under this
conditional density for $y_{it}$.

The approach now becomes similar to the univariate method of Section
\ref{sec:vol}. The marginal score function is
\[
g(y_{it};\alpha_{it},\theta)=\frac{1}{2}y_{it}^{2}e^{-\alpha_{it}}-\frac{1}%
{2}=\frac{1}{2}(\varepsilon_{it}^{2}-1).
\]
The distribution function for the function $g$ under $f(y_{it}\mid\alpha
_{it},\mathcal{F}_{t-1};\theta)$ is
\[
F_{g}\left\{  g(y_{it};\alpha_{it},\theta)\mid\alpha_{it};\theta\right\}
=F_{\chi_{1}^{2}}(\varepsilon_{it}^{2}),
\]
where $F_{\chi_{1}^{2}}$ is the distribution function for the chi-square
distribution with one degree of freedom. This is similar to Section
\ref{sec:vol}. We denote $u_{t}=(u_{1t},...,u_{Nt})^{\prime}$ where
$u_{it}=F_{\chi_{1}^{2}}(\varepsilon_{it}^{2})$. It is clear that $u_{t}$ is
distributed according to a copula, according to the definition given in
Section \ref{sec:copula}.

The innovation term $\eta_{t}$, an $N\times1$ vector, is derived as
\begin{equation}
\eta_{it}=\Phi^{-1}\left(  u_{it}\right)  =\Phi^{-1}\left[  F_{\chi_{1}^{2}%
}(\varepsilon_{it}^{2})\right]  .\label{eq:vol_innov_mult}%
\end{equation}
The resulting iGASC\ models is clearly observation-driven as the multivariate
prediction density $f(y_{t}|y_{1:t-1};\theta)$ can be calculated. In this case
$\theta$ represents all the $N(N-1)/2$ elements of $\Sigma$ and the parameters
$(\mu_{i},\phi_{i},\psi_{i})^{\prime}$ for $i=1,...,N$.

There is a caveat to note in the formation of the innovations vector $\eta
_{t}$. Whilst each element $\eta_{it}$ will be marginally standard Gaussian,
the vector will $\eta_{t}$ not be multivariate Gaussian. This is a consequence
of the copula construction as the terms $u_{it}$ in (\ref{eq:vol_innov_mult})
are marginally uniform but the vector $u_{t}$ is not a multivariate uniform
random variate. For the multivariate iGASc model the marginal and predictive
distributions of $\alpha_{it}$ remain simple to calculate and have a form
analogous to those of Sections \ref{sec:properties} and \ref{sec:forecasting}.
However, the multivariate forecast distribution for the vector $\alpha_{t}$,
for more than one period ahead, is unavailable in closed form. 

\subsection{Multivariate results\label{sec:mult_results}}

\begin{table}[ptb]
\centering{%
\begin{tabular}
[c]{cccc}\hline
\multicolumn{4}{c}{Trivariate iGASC-Gaussian}\\\hline
Nikkei & DAX & Hang Seng & Correlation\\\hline
\textit{$\mu_{1}$} & \textit{$\mu_{2}$} & \textit{$\mu_{3}$} & \textit{$\rho
_{N,D}$}\\
0.03874 & 0.01744 & 0.01402 & 0.29456\\
(0.02550, 0.05198) & (0.00996, 0.02492) & (0.00807, 0.01997) & (0.25733,
0.33179)\\
\textit{$\phi_{1}$} & \textit{$\phi_{2}$} & \textit{$\phi_{3}$} &
\textit{$\rho_{D, HS}$}\\
0.94421 & 0.97053 & 0.97193 & 0.35937\\
(0.92430, 0.96412) & (0.95703, 0.98403) & (0.95977, 0.98409) & (0.32367,
0.39507)\\
\textit{$\psi_{1}$} & \textit{$\psi_{2}$} & \textit{$\psi_{3}$} &
\textit{$\rho_{N, HS}$}\\
0.15444 & 0.10583 & 0.08174 & 0.52086\\
(0.00996, 0.02492) & (0.07856, 0.13310) & (0.06166, 0.10182) & (0.49110,
0.55062)\\\hline
\end{tabular}
}\caption{\textsl{Parameter estimates and 95}$\%$\textsl{ confidence intervals
for trivariate iGASC model applied to daily returns for the Nikkei, DAX and
Hang Seng indices. The distribution of the standardised return }%
$\varepsilon_{t}$\textsl{ is considered as multivariate Gaussian. }}%
\label{tab:tri_param_daily}%
\end{table}\begin{table}[ptb]
\centering{%
\begin{tabular}
[c]{cccc}\hline
\multicolumn{4}{c}{Trivariate iGASC-Gaussian}\\\hline
Nikkei & DAX & Hang Seng & Correltation\\\hline
\textit{$\mu_{1}$} & \textit{$\mu_{2}$} & \textit{$\mu_{3}$} & \textit{$\rho
_{N, D}$}\\
0.27965 & 0.14609 & 0.51559 & 0.59429\\
(-0.00149, 0.56079) & (0.02917, 0.26301) & (-3.17686, 4.20804) & (0.53818,
0.65040)\\
\textit{$\phi_{1}$} & \textit{$\phi_{2}$} & \textit{$\phi_{3}$} &
\textit{$\rho_{D, HS}$}\\
0.86064 & 0.92443 & 0.71730 & 0.52995\\
( 0.71877, 1.00251) & (0.86295, 0.98591) & (-1.29089, 2.72549) & (0.45458,
0.60532)\\
\textit{$\psi_{1}$} & \textit{$\psi_{2}$} & \textit{$\psi_{3}$} &
\textit{$\rho_{N, HS}$}\\
0.15184 & 0.11469 & 0.13833 & 0.58844\\
(0.06820, 0.23548) & (0.05573, 0.17365) & (-0.29974, 0.57640) & (0.52842,
0.64846)\\\hline
\end{tabular}
}\caption{\textsl{Parameter estimates and 95}$\%$\textsl{ confidence intervals
for trivariate iGASC model applied to weekly Nikkei returns, weekly DAX
returns and weekly Hang Seng returns. The distribution of the standardised
return }$\varepsilon_{t}$\textsl{ is considered as multivariate Gaussian. }}%
\label{tab:tri_param_weekly}%
\end{table}\begin{figure}[th]
\centering
\includegraphics[width=0.9\textwidth]{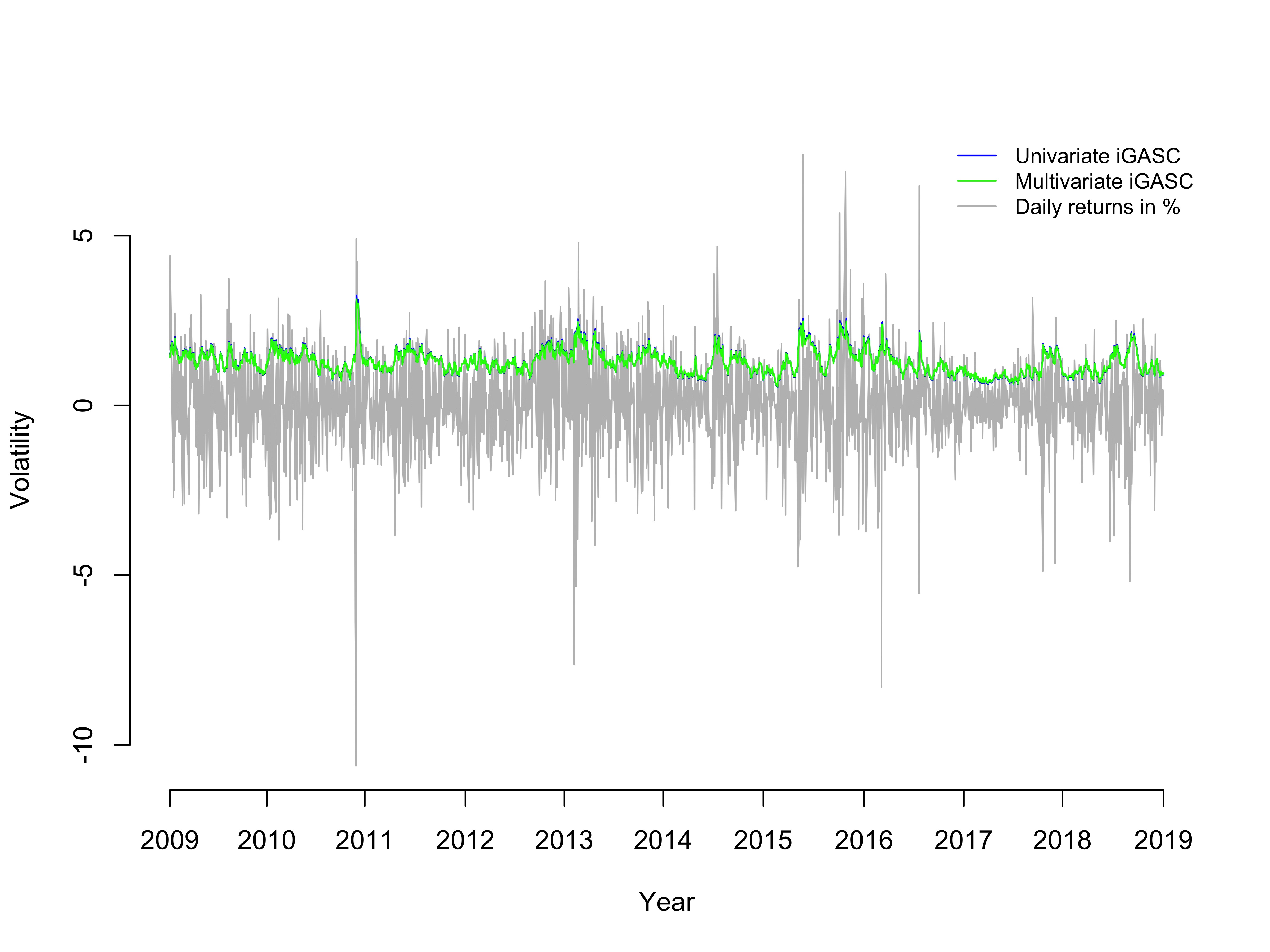}\caption{\textsl{Results
for the daily continuously compounded percentage returns for the Nikkei index.
The standard deviations }$\sigma_{t}=exp(\alpha_{t}/2)$\textsl{ for the
iGASC-Gaussian univariate and trivariate model.}}%
\label{fig_daily_vol_nikk_comp}%
\end{figure}\begin{figure}[th]
\centering\includegraphics[width=0.9\textwidth]{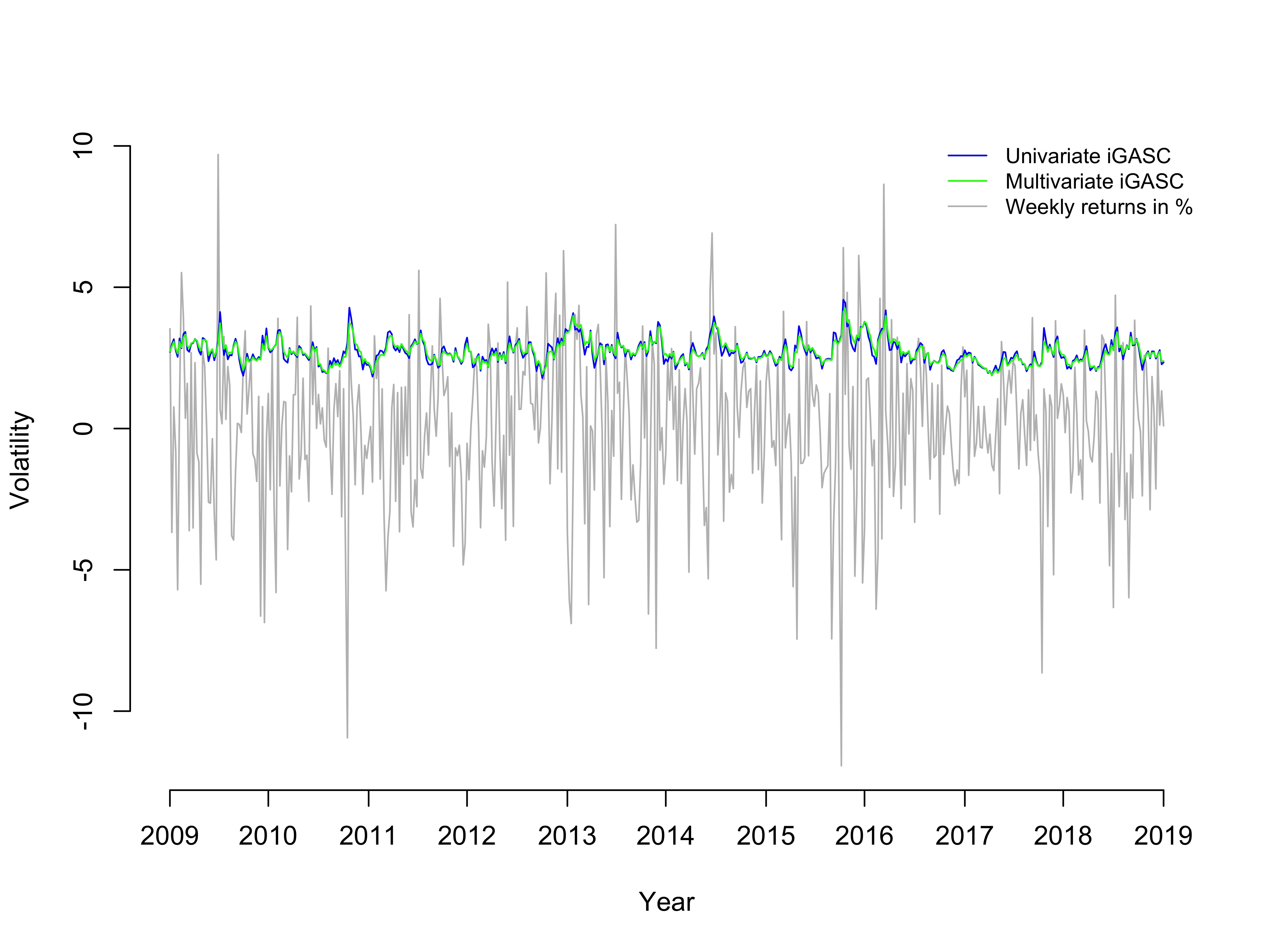}\caption{\textsl{Results
for the 516 weekly continuously compounded percentage returns for the Nikkei.
The standard deviations }$\sigma_{t}=exp(\alpha_{t}/2)$\textsl{ for the
iGASC-Gaussian univariate and trivariate model.}}%
\label{fig_weekly_vol_nikk_comp}%
\end{figure}

The multivariate model of (\ref{eq:mult_sv}) is applied to three daily and
weekly stock index returns. As in Section \ref{sec:ret_data}, continuously
compounded returns in percentages are analysed. The three series considered
are the Nikkei, DAX and Hang Seng indices. The data is again from 2009-04-30
to 2019-04-30. We denote the elements of the correlation matrix as
\[
\boldsymbol{\Sigma}=%
\begin{pmatrix}
1 & \rho_{N,D} & \rho_{D,HS}\\
\rho_{N,D} & 1 & \rho_{N,HS}\\
\rho_{D,HS} & \rho_{N,HS} & 1
\end{pmatrix}
,
\]
where $\rho_{N,D}$ is the conditional correlation between Nikkei and Hang Seng
indices, $\rho_{D,HS}$ the correlation between the DAX and Hang Seng indices
and $\rho_{N,HS}$ the correlation between the Nikkei and Hang Seng indices.
Maximum likelihood is employed to estimate the parameters. This is extremely
simple to implement and runs very quickly.

The parameter estimates from the maximum likelihood procedure applied to the
daily data are provided in Table \ref{tab:tri_param_daily}. The estimates for
$(\mu_{1},\phi_{1},\psi_{1})^{\prime}$ for the Nikkei daily returns (first
column) are quite similar to those in Table \ref{tab:param_nikkei} as would be
expected. The estimates of the three elements of $\boldsymbol{\Sigma}$ (final
column) are positive and clearly far from zero, indicating that modelling the
correlation is important. The persistence parameters for the DAX and Hang Seng
indices, $\phi_{2}$ and $\phi_{3}$, are considerably higher than for the
Nikkei index. The results for weekly data are provided in Table
\ref{tab:tri_param_weekly}. The multivariate iGASc model of (\ref{eq:mult_sv})
implies the marginal univariate models of Section \ref{sec:vol}. The parameter
estimates will not be the same as the information in the likelihood will be
richer for the multivariate models. However, there should be close agreement
and this can be seen in Figure \ref{fig_daily_vol_nikk_comp}. The conditional
standard deviation for the daily Nikkei index is plotted for both the
univariate model and for the multivariate model. It can be seen that there is
very close correspondence between the two series. This closeness is also seen
for the weekly data in Figure \ref{fig_weekly_vol_nikk_comp}.

%Please add the following required packages to your document preamble:
%\usepackage{graphicx}

\section{Discussion\label{sec:discussion}}

The paper shows a new approach for the modelling of dynamic time series. The
resulting models inherit many of the attractive properties associated with
latent parameter driven models. However, in contrast to parameter driven
models, the one step ahead prediction density is available enabling maximum
likelihood to be performed straightforwardly. The approach relies on the
probability integral transform method, applied to a function of interest, for
univariate series. For multivariate evolving parameters, this extends
naturally to consideration of a copula method applied to the function of
interest. We have deliberately restricted the function of interest to the
conditional score and have seen that this results in sensible properties. This
ties into the related literature on generalized autoregressive score (GAS)
modelling. We have shown how to construct these iGASc models for both dynamic
volatility and duration models. The methods appear competitive with existing
approaches, for example GARCH\ models, whilst having tractable marginal and
predictive densities. The multivariate extension of the models has been
considered for modelling the volatility of several returns. This leads to a
simple model which is applied to three stock indices. This new approach can be
easily extended. A very broad class of dynamic models can be considered by
adopting the modelling strategy. Additionally, there are many ways in which
the copula function of interest can be constructed for multivariate modelling.
We have chosen to use the marginal conditional score function. However, in
many contexts it may be the full conditional score is more appropriate. An
open questions remains when the number of observations, at each time period,
is either greater than or less than the number of evolving states at each time
period. 

\newpage

\section*{Acknowledgements}

%\clearpage
%\section*{References}
\bibliographystyle{plain}
\bibliography{reference}

%\clearpage
%\appendix

%\section{}
%\label{appx:1}

\end{document}